\documentclass[a4paper,11pt,notitlepage]{article}
\usepackage{multirow,url,color}
\usepackage[utf8x]{inputenc}

\title{\textbf{NIEL Dose Analysis on triple and single junction InGaP/GaAs/Ge solar cells irradiated with electrons, protons and neutrons}}
\author{Roberta Campesato$^{1}$, Carsten Baur$^{2}$, Mario Carta$^{3}$, Mariacristina Casale$^{1}$, Davide Chiesa$^{4,5}$,\\ Massimo Gervasi$^{4,5}$, Enos Gombia$^{6}$, Erminio Greco$^{1}$, Aldo Kingma$^{6}$, Massimiliano Nastasi$^{4,5}$,\\ Ezio Previtali$^{4,5}$, Pier Giorgio  Rancoita$^{4}$, Davide Rozza$^{4,5}$, Emilio Santoro$^{3}$, Mauro Tacconi$^{4,5}$.}
\date{}

\usepackage[utf8x]{inputenc}
\usepackage{amsmath}
\usepackage{amsfonts}
\usepackage{amssymb}
\usepackage{graphicx}
\usepackage[top=2.3cm,bottom=2.3cm,left=2.3cm,right=2.3cm]{geometry}

\begin{document}
\maketitle

\begin{center}
\scriptsize{
$^1$ \textit{CESI, via Rubattino 54, I-20134 Milan, Italy}\\
$^2$ \textit{ESA/ ESTEC, Keplerlaan 1, 2201 AZ Noordwijk, The Netherlands}\\
$^3$ \textit{ENEA C.R. CASACCIA - FSN-FISS-RNR - S.P.040 via Anguillarese 301, 00123 S. Maria di Galeria (Roma) Italy}\\
$^3$ \textit{INFN Sezione di Milano Bicocca, I-20126 Milan, Italy}\\
$^5$ \textit{Universit\`a di Milano Bicocca, I-20126 Milan, Italy} \\
$^6$ \textit{IMEM-CNR Institute, Parco Area delle Scienze 37/A, 43124 Parma, Italy}\\
}
\end{center}

\begin{abstract}
Triple junction (InGaP/GaAs/Ge) and single junction (SJ) solar cells were irradiated with electrons, protons and neutrons. The degradation of remaining factors was analyzed as function of the induced Displacement Damage Dose (DDD) calculated by means of the SR-NIEL (Screened Relativistic Non Ionizing\ Energy Loss) approach.  In particular, the aim of this work is to analyze the variation of the solar cells remaining factors due to neutron irradiation with respect to those previously obtained with electrons and protons. The current analysis confirms that the degradation of the $P_{max}$ electrical parameter is related by means of the usual semi-empirical expression to the displacement dose, independently of type of the incoming particle. $I_{sc}$ and $V_{oc}$ parameters were also measured as a function of the displacement damage dose. Furthermore, a DLTS analysis was carried out on diodes - with the same epitaxial structure as the middle sub-cell - irradiated with neutrons.
\end{abstract}

\begin{center}
to appear in the Proceedings of the\\ 46th IEEE PHOTOVOLTAIC SPECIALISTS CONFERENCE (PVSC 46), \\ June 16-21 (2019), Chicago (USA)

\end{center}

\section{Introduction}
The prediction of solar cell degradation, due to radiation present in space environment, is of primary importance in the preparation of space missions towards harsh radiation orbits. The radiation analysis of solar cells is required to predict the End Of Life (EOL) performances of the solar arrays. Degradation of solar cell electrical performances depends on the energy, the fluence and the type of the incident particles (electrons, protons, neutrons, etc.).
\par Two methodologies are currently adopted by the space actors to perform on-orbit solar cell performance predictions: the Equivalent Fluence method developed by JPL (Jet Propulsion Laboratory) \cite{Anspaugh} and the Displacement Damage Dose (DDD) method developed by NRL (Naval Research Laboratory) \cite{Messengers}. Even if the first one is the mainly used because of its heritage, it has the disadvantage of requiring a large number of irradiation tests with different particles having various energies and fluences. On the contrary, the more recent DDD approach consents to predict the EOL behavior of solar cells starting from a reduced number of irradiation tests allowing for a rapid analysis of emerging cells technologies. The key aspect of this method is that it is based on the calculation of the NIEL (Non Ionizing Energy Loss) doses which, in turn, depends on the amount of the permanent displacement damage induced by particle interactions inside the device and resulting in the degradation of its electrical performance. Following this approach, a single characteristic degradation curve can be obtained regardless of the particle type when the degradation is related to the actual values of the displacement damage dose only.
\par In the present work, the results of electron, proton and neutron irradiation tests, performed on triple junction solar cells and related isotype sub-cells manufactured by CESI, will be presented. The NIEL doses, which depend on the displacement threshold energy $E_{d}$, were obtained by means of the SR-NIEL tool (\cite{Boschini} and Chapters 2, 7 and 11 in \cite{Leroy}). The degradation curve for each cell type, is obtained by plotting the remaining factors (the ratio of the end-of-life EOL value to the beginning-of-life BOL value) of a given electrical parameter as a function of the calculated DDD.

\section{Description of irradiated samples}
InGaP/InGaAs/Ge\ TJ solar cells and related component cells  with AM0 efficiency class 30\%  (CTJ30), have been manufactured as $2\times2$ cm$^{2}$ solar cells and 0.5 mm diameter diodes (only top and middle sub cell) \cite{Gori}.
\begin{figure}[h!]
\centering
\includegraphics[width=0.9\textwidth]{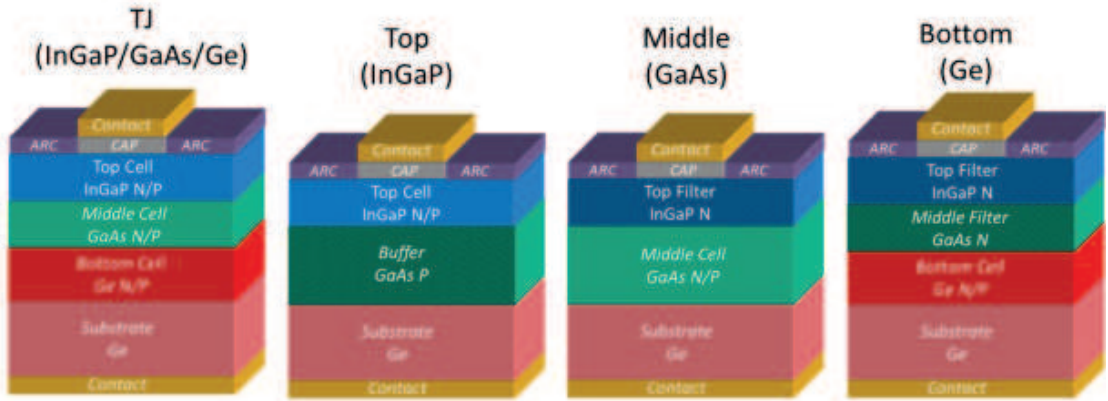}
\caption{Schema of\  triple junction (TJ) and single junction (SJ) isotype sub-cells (Top, Middle and Bottom).}
\label{fig:1}
\end{figure}
The basic structure of the solar cells is reported in Fig. \ref{fig:1}. The TJ solar cell is composed by a germanium bottom junction obtained by diffusion into the germanium P-type substrate, a middle junction of (In)GaAs, whose energy gap is around 1.38 eV and a top junction of InGaP with an energy gap of 1.85 eV. Component cells are single-junction (SJ) cells which shall be an electrical and optical representation of the subcells inside the TJ cell. Therefore, to manufacture them, special attention was put to reproduce the optical thicknesses of all the upper layers present in the TJ structure. Top and middle sub cells for DLTS analysis were also manufactured as diodes with 0.5 mm diameter using a mesa etch to remove the edge defects related to cutting.

\section{Experimental Irradiation Procedure}
Solar cells have been measured in BOL conditions and then after irradiation, allowing a self-annealing duration of about one month for electrons and protons and about two months for neutrons.
\par TJ solar cells and component cells have been irradiated with protons and electrons at different energies and fluences (see Table \ref{Tab1}1) \cite{Campesato1,Campesato2}.
\par The main electrical parameters were recorded and the ratio EOL/BOL (known as Remaining Factor) was calculated. The bottom junction is highly degraded after proton irradiation whereas it is highly radiation resistant when irradiated with electrons.
\begin{table}[h!]
\centering
\caption{Electron and proton irradiation test conditions.}
\begin{tabular}{|c|c|c|c|}
\hline
\multicolumn{2}{|c|}{Protons Irradiated at CSNSM Orsay (France)} & \multicolumn{2}{|c|}{Electrons Irradiated at Delft (The Netherlands)} \\
\hline
\multicolumn{1}{|c|}{\textbf{Energy [MeV]}} &
\multicolumn{1}{|c|}{\textbf{Fluence [p+/cm$^2$]}} &
\multicolumn{1}{|c|}{\textbf{Energy [MeV]}} &
\multicolumn{1}{|c|}{\textbf{Fluence [e-/cm$^2$]}}\\
\hline
\multicolumn{1}{|c|}{\multirow{1}{*}{\begin{tabular}{c}0.7\\\end{tabular}}} &
\multicolumn{1}{|c|}{1.7E+11} &
\multicolumn{1}{|c|}{\multirow{1}{*}{\begin{tabular}{c}1\\\end{tabular}}} &
\multicolumn{1}{|c|}{1.0E+14} \\
\multicolumn{1}{|c|}{} &
\multicolumn{1}{|c|}{3.3E+11} &
\multicolumn{1}{|c|}{} &
\multicolumn{1}{|c|}{5.0E+14} \\
\multicolumn{1}{|c|}{} &
\multicolumn{1}{|c|}{4.5E+11} &
\multicolumn{1}{|c|}{} &
\multicolumn{1}{|c|}{1.0E+15} \\
\hline
\multicolumn{1}{|c}{\multirow{1}{*}{\begin{tabular}{c}1\\\end{tabular}}} &
\multicolumn{1}{|c}{4.5E+10} &
\multicolumn{1}{|c}{\multirow{1}{*}{\begin{tabular}{c}1.5\\\end{tabular}}} &
\multicolumn{1}{|c|}{5.0E+14} \\
\multicolumn{1}{|c}{} &
\multicolumn{1}{|c}{2.3E+11} &
\multicolumn{1}{|c}{} &
\multicolumn{1}{|c|}{1.0E+15} \\
\hline
\multicolumn{1}{|c}{\multirow{1}{*}{\begin{tabular}{c}2\\\end{tabular}}} &
\multicolumn{1}{|c}{4.2E+11} &
\multicolumn{1}{|c}{\multirow{1}{*}{\begin{tabular}{c}3\\\end{tabular}}} &
\multicolumn{1}{|c|}{2.0E+14} \\
\multicolumn{1}{|c}{} &
\multicolumn{1}{|c}{8.4E+11} &
\multicolumn{1}{|c}{} &
\multicolumn{1}{|c|}{4.0E+14} \\
\hline
\end{tabular}\label{Tab1}
\end{table}
The neutron irradiations were performed at the TRIGA reactor in Casaccia (Rome). The neutron spectral fluence of that reactor was determined \cite{ASIF} using experimental data of activation rates and corresponding cross sections of various elements \cite{Chiesa1,Chiesa2}. The solar cells were irradiated together with monitor samples to get a better than 5 percent accuracy on neutron fluence. Spatial characterization was performed by means of Al-Co monitor sample \cite{Chiesa3}. The analysis of $P_{max}$ versus DDD is reported in the next section.

\section{NIEL Analysis}
The photovoltaic parameters of the TJ and SJ cells are investigated by plotting the remaining factors of each electrical parameter as a function of displacement damage dose (DDD).
\par For electrons and protons, DDD, expressed in units of MeV/g, is computed from:
\begin{equation}\label{Eq1}
DDD=\Phi\frac{dE_{de}}{d\chi}
\end{equation}
where $\Phi$ is the fluence in cm$^{-2}$ of traversing particles and $\frac{dE_{de}}{d \chi}$ is the so-called non-ionizing energy loss NIEL which expresses the amount of energy deposited by an incident particle passing through a material and resulting in displacement processes. This quantity, expressed in units of MeV cm$^{2}$/g, is obtained by means of the SR (Screened Relativist) treatment \cite{Boschini,Leroy}
\begin{equation}\label{Eq2}
\frac{dE_{de}}{d \chi }=\frac{N}{A} \int _{E_{d}}^{E_{R}^{max}}P_k\left(E_{R}\right)\frac{d \sigma \left( E,E_{R} \right) }{dE_{R}}dE_{R}
\end{equation}
where $\chi$  is the absorber thickness in g/cm$^{2}$, $N$ is the Avogadro constant, $A$ is the atomic weight of the medium; $E$ is the kinetic energy of the incoming particle; $E_{d}$ and $E_{R}^{max}$ are displacement threshold energy and the maximum energy transferred to the recoil nucleus respectively; $P_k(E_{R})$ is the partition energy for the recoil nucleus (i.e. the part of the recoil energy deposited in displacements); $d\sigma(E,E_{R})/dE_{R}$ is the differential cross section for any incoming particle with kinetic energy $E$ resulting in a nuclear recoil with kinetic energy $E_{R}$.
\par It should be remarked that the energy of incoming protons has to be modified as a function of the depth into the device, according to the amount of energy deposited which can be calculated (by means of SRIM \cite{Ziegler}) from the stopping power in each absorbing layer \cite{Campesato1,Campesato2}. In this work, the TJ solar cell was approximated by a GaAs cell single junction cell since the middle cell is the one that mainly affects the overall TJ performances at EOL.
\par For fast neutrons from nuclear reactors, the displacement damage dose can be computed (see \cite{Boschini} and Chapter 7 in \cite{Leroy}) from:
\begin{equation}\label{Eq3}
 DDD=\frac{N}{A} \int_{E_{min}}^{E_{max}}D\left(E\right)\phi\left(E\right)dE
\end{equation}
where $D(E)$ is the \textit{damage function} (also referred to as \textit{displacement kerma function}) in units of MeV cm$^{2}$ and $\phi(E)$ is the neutron spectral fluence in n cm$^{-2}$ MeV$^{-1}$. The damage functions were obtained from the SR-NIEL calculator available in \cite{Boschini} based on the NJOY code \cite{MacFarlane}. For the present calculation the spectral fluence used is the one for the TRIGA reactor (see Section III).
\par The following semi-empirical equation is used to fit the experimental remaining factors ($RF_{par}$) of each electrical parameter as a function of DDD:
\begin{equation}\label{Eq4}
 \frac{RF_{par}\left( EOL \right)}{RF_{par}\left( BOL \right)}= \left( 1-A \right) -C \cdot log_{10} \left[ 1+\frac{DDD \left( E_{d} \right) }{DDD_{x}} \right]
 \end{equation}
being A, C and DDD$_{x}$ fit parameters. It should also be noted that $A$ is only relevant for bottom cells.
\par The displacement threshold energies, $E_{d}$, were found using a routine which globally minimizes the square root relative difference (SRRD) of the points with respect to the fitted curve.
\par
\begin{figure}[h!]
\centering
\includegraphics[width=0.9\textwidth]{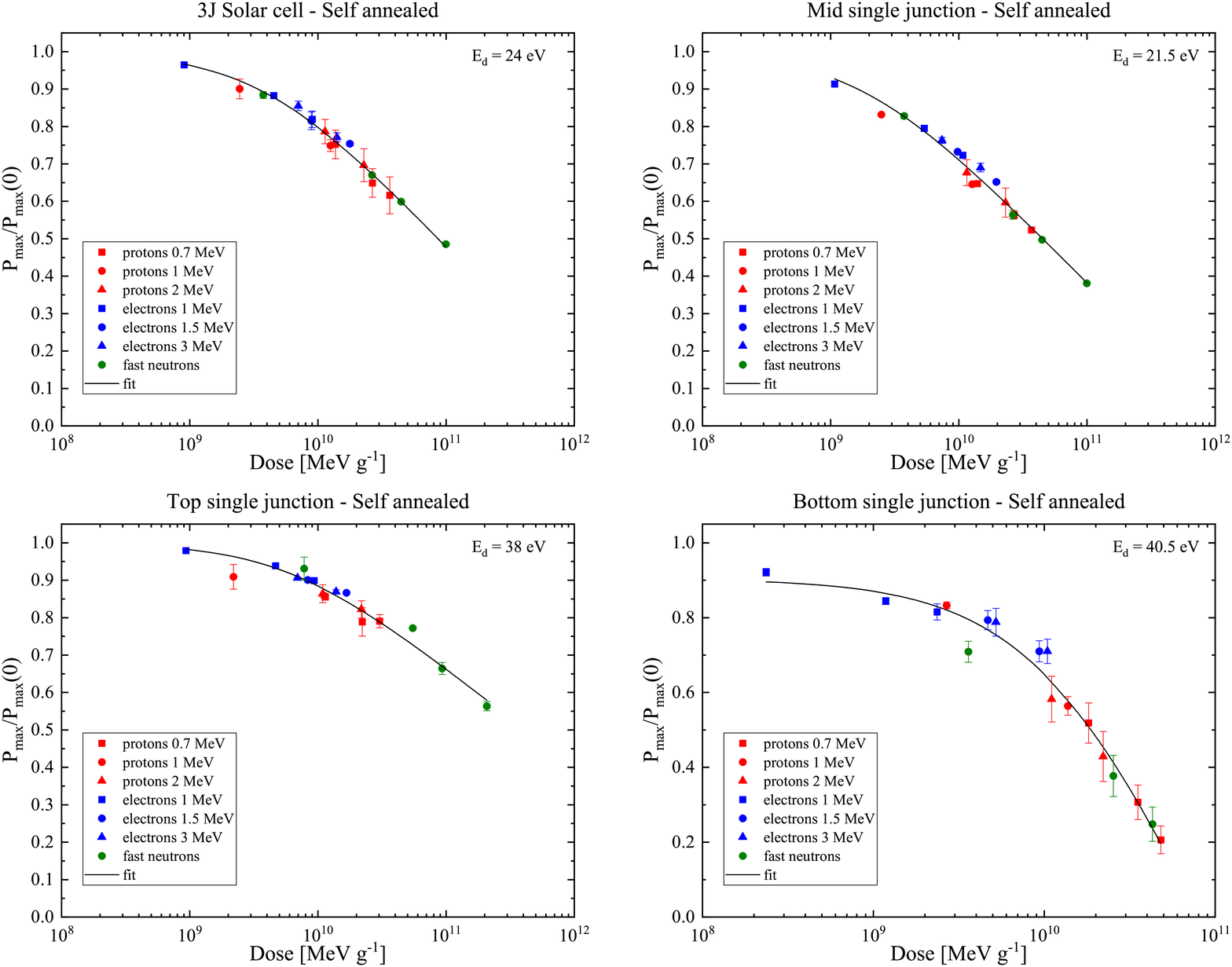}
\caption{Optimal fit of $P_{max}$ degradation curve for TJ, mid cell, top cell and bottom cell.}
\label{fig:2}
\end{figure}

 Fig. \ref{fig:2} shows the remaining power factor as function of the DDD; the values of $E_{d}$ in eV indicated are those found by the global minimization with respect to the semi-empirical expression \ref{Eq4}. It is worth to remark that, for neutrons, the calculated NIEL is almost independent on the values of $E_{d}$.
\par By inspection of Fig. \ref{fig:2} one can remark the good agreement of the degradation of remaining power factor as a function of the DDD with respect to Eq. \ref{Eq4}, independently of the type of incoming particle.

\section{DLTS Analysis}
\begin{figure}[h!]
\centering
\includegraphics[width=0.75\textwidth]{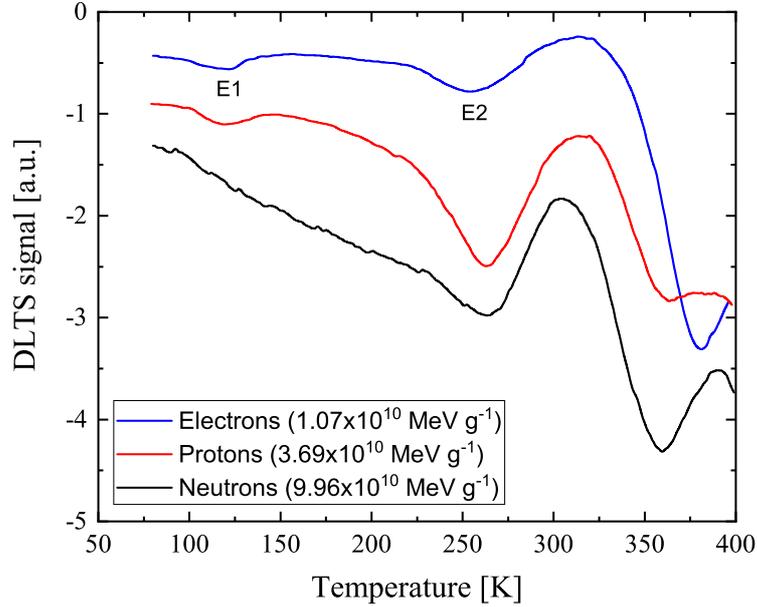}
\caption{Comparison of the DLTS spectra of middle junctions irradiated by protons ($3.69\times10^{10}$ MeV g$^{-1}$), electrons ($1.07\times10^{10}$ MeV g$^{-1}$) and neutrons ($9.96\times10^{10}$ MeV g$^{-1}$), respectively. Emission rate $=46$ s$^{-1}$, pulse width=$500\ \mu$s, reverse voltage $V_r=-1.5$ V, pulse voltage $V1=-0.1$ V.}
\label{fig:3}
\end{figure}

A DLTS investigation regarding deep levels, induced by neutron irradiation, was carried out using samples described in \cite{Campesato1,Campesato2}. In those articles the experimental procedure and technique were also discussed.  In particular, diodes of 0.5 mm in diameter were prepared using the same epitaxial structure of the middle sub-cell. The DLTS spectra - obtained from middle-cell diodes irradiated with electrons, protons and neutrons - are shown in Fig. \ref{fig:3}.
By inspection of Fig. \ref{fig:3}, one can observe how the ratio of the peak amplitudes E2/E1 is much larger for the proton irradiated sample than for the electron irradiated one \cite{Campesato1,Campesato2}. In addition, the peak E1 was detected a) at all doses for samples irradiated with electrons, b) only at the highest doses for those irradiated with protons and, c) finally, it was not untangled from the background in samples irradiated with neutrons.  The E1 and E2 deep levels were identified as majority carrier traps in the p-type bulk bases \cite{Campesato1,Campesato2}.
\begin{figure}[h!]
\centering
\includegraphics[width=0.9\textwidth]{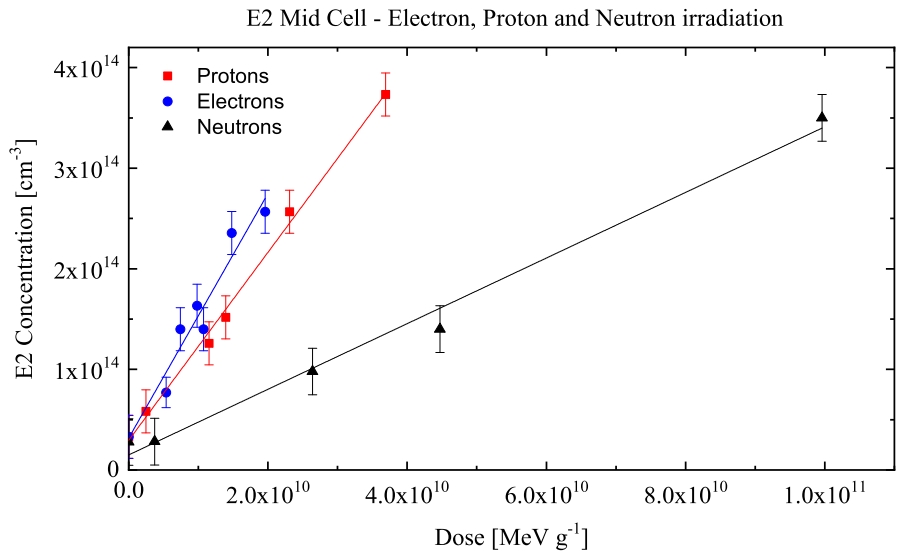}
\caption{Concentration of E2 traps induced by irradiations with electrons, protons and neutrons in middle sub cell diodes as a function of DDD with $E_d=21.5$ eV.}
\label{fig:4}
\end{figure}
\par The DLTS peak heights measured allowed for determining the E2 defect concentrations - introduced by displacement damage - which are shown in Fig. \ref{fig:4} as function of DDD (the data for concentrations obtained with electron and proton irradiations are from \cite{Campesato1,Campesato2}). The concentrations of E2 level were found to depend almost linearly on NIEL dose, although with a slope which depends on the type of the damaging particle. The interpretation of such a behaviour and how it relates with the correlation exhibited in Fig. \ref{fig:2} deserve further investigations.

\section{Conclusions}
TJ InGaP/GaAs/Ge solar cells and related component cells, manufactured by CESI, were irradiated with electrons and protons and, finally, with neutrons at the TRIGA reactor in Casaccia. Solar cell electrical performances degradation was analyzed as a function of the induced Displacement Damage Dose (DDD) computed following the SR-NIEL approach.
\par The experimentally obtained remaining power factors are well represented by a single semi-empirical expression as a function of the DDD. In addition, the concentration of the E2 deep level was found to depend almost linearly on DDD with a slope which, in turn, depends on the particle type.
\par The current results make the usage of neutron particles a possible candidate for testing solar cell degradation for application in space environment.

\vspace*{0.5cm}
\footnotesize{{\bf Acknowledgment:}{Part of this work - including electron and proton irradiations - was supported by ESA contract 4000116146/16/NL/HK with title ``Non-Ionizing Energy Loss (NIEL) Calculation and Verification''. Furthermore, the fast neutron irradiation of solar cells and the neutral fluence determination was accomplished under ASI-ENEA ASIF implementation agreement no. 2017-22-H.0 and ASI-INFN ASIF implementation agreement no. 2017-15-H.0.}}

\end{document}